%% file: OPTICAmeetings.tex
\documentclass[letterpaper,10pt]{article} 

\usepackage{opticameet3} 

\usepackage{amsmath,amssymb}
\usepackage{pgfplotstable}
\usepackage{pgfplots}
\usepackage{pgfplots}
\usepackage{subcaption}
\usepackage{siunitx}
\usepackage{url}
\usepackage[utf8]{inputenc}
\usepackage{textcomp}
\usepackage{newunicodechar}
\newunicodechar{−}{\ensuremath{-}}
\usepackage[colorlinks=true,bookmarks=false,citecolor=blue,urlcolor=blue]{hyperref}
\usepackage[capitalize]{cleveref}
\usetikzlibrary{arrows.meta,calc,decorations.markings,math}
\usepackage{cite}
\usepackage[font=small]{caption}
\usetikzlibrary{backgrounds}
\usepackage{import}

\usepackage{pgfplots}
\usepackage{titlesec}
\usepgfplotslibrary{fillbetween}
\usepgfplotslibrary{groupplots}
\pgfplotscreateplotcyclelist{foo}{
        {C1,mark=*},
        {C2,mark=square*},
        {C3, mark = triangle*},
        {C4,mark=+},
        {C5, mark = diamond*},
        {C6, mark = x}
      }

\newcommand\authormark[1]{\textsuperscript{#1}}

\titlespacing{\section}{0pt}{10pt}{5pt}
\usepackage{amsmath,amssymb} 
\definecolor{C0}{RGB}{031, 119, 180} 
\definecolor{C1}{RGB}{255, 127, 014} 
\definecolor{C2}{RGB}{044, 160, 044} 
\definecolor{C3}{RGB}{215, 039, 040} 
\definecolor{C4}{RGB}{148, 103, 189} 
\definecolor{C5}{RGB}{140, 086, 075} 
\definecolor{C6}{RGB}{227, 119, 194} 
\definecolor{C7}{RGB}{127, 127, 127} 
\definecolor{C8}{RGB}{188, 189, 034} 
\definecolor{C9}{RGB}{023, 190, 207} 
\definecolor{C0l}{RGB}{174, 199, 232} 
\definecolor{C1l}{RGB}{255, 187, 120} 
\definecolor{C2l}{RGB}{152, 223, 138} 
\definecolor{C3l}{RGB}{255, 152, 150} 
\definecolor{C4l}{RGB}{197, 176, 213} 
\definecolor{C5l}{RGB}{196, 156, 148} 
\definecolor{C6l}{RGB}{247, 182, 210} 
\definecolor{C7l}{RGB}{199, 199, 199} 
\definecolor{C8l}{RGB}{219, 219, 141} 
\definecolor{C9l}{RGB}{158, 218, 229} 

\usepackage{graphicx}
\usepackage{subcaption}
\newlength{\myimagewidth}

\begin{document}
\setlength{\myimagewidth}{\dimexpr\columnwidth/3-1em\relax}

\title{Capacity and SKR Trade-off in Coexisting Classical and CV-QKD Metro-Reach Links}

\vspace{-6mm}
\author{Çağla~Özkan\authormark{1,*}, Lucas~Alves~Zischler\authormark{2}, Kadir~Gümüş\authormark{1},
João dos Reis Frazão\authormark{1},
Cristian~Antonelli\authormark{2}, and Chigo~Okonkwo\authormark{1}}

\vspace{-1mm}
\address{\authormark{1}High-capacity Optical Transmission Laboratory, Eindhoven University of Technology, The Netherlands\\
\authormark{2}Department of Physical and Chemical Sciences, University of L’Aquila, L’Aquila, Italy}

\email{\authormark{*}c.o.ozkan@tue.nl} 

\vspace{-7.8mm}
\begin{abstract}
We evaluate the coexistence impairments of CV-QKD in metropolitan DWDM networks. Using a comprehensive interference model, we develop guardband rules where band-edge placement with 100--150 GHz guardbands doubles SKR with minimal classical capacity loss.

\end{abstract}
\vspace{-0.5mm}
\section{Introduction}
\vspace{-2mm}
Quantum key distribution (QKD) provides information-theoretically secure key generation. While CV and DV-QKD protocols are proven over metropolitan distances~\cite{Pirandola_2020}, most DV implementations require dedicated dark fibres or band separation, increasing costs and limiting scalability. CV-QKD is inherently compatible with coherent detection and benefits from a higher noise tolerance~\cite{Qi_2010}, allowing coexistence with internet traffic over the same dense wavelength division multiplexing (DWDM) fibre~\cite{Eriksson2019,hajomer2025coexistencecontinuousvariablequantumkey}. In such hybrid links, spontaneous Raman scattering (SpRS) and four-wave mixing (FWM) from classical channels degrade performance; SpRS dominates at long distances, while FWM, with its cubic power scaling, becomes a dominant impairment in 10–25~km DWDM systems~\cite{Kumar_2015}. For modelling, previous work has optimised wavelength allocation and quantified Raman and FWM-limited performance for QKD-secured DWDM networks~\cite{Niu:18}, but systematic, power-dependent guardband design rules and quantitative SKR–capacity trade-offs for CV-QKD coexistence have not been explored in detail. Using the coexistence model of~\cite{zischler2025accurateeffectivemodelcoexistence}, this work shows that by sweeping the guardband size, launch power and transmission distance, the quantum channel placed at the edge of the C-band with 100–150~GHz guardbands can double the SKR while causing only a $3.4\%$ loss in classical capacity.
\vspace{-2mm}
\section{Analytical Model and Framework}
\vspace{-2mm}
We consider a C-band link where $N_{\mathrm{cl}}$ classical WDM channels and one CV‑QKD channel co‑propagate over a standard single‑mode fibre. Classical channels at wavelengths $\lambda_i$ are launched with powers $P_i$, while the quantum channel at $\lambda_q$ experiences linear loss $T = 10^{-\alpha L/10}$  (with the $\alpha$ attenuation coefficient and $L$ the fibre length) and nonlinear interference generated by classical traffic. The coexistence model in~\cite{zischler2025accurateeffectivemodelcoexistence} gives the nonlinear interference power $P_{\mathrm{int}}$ at $\lambda_q$, accounting for SpRS, FWM and other out‑of‑band effects with full frequency dependence. In this work, $P_{\mathrm{int}}$ is used as the coexistence‑induced interference at the quantum wavelength and mapped to input‑referred excess noise for a coherent CV‑QKD link via the standard photon‑number relation~\cite{Qi_2010}: the mean number of noise photons per spatio-temporal mode generated from the classical signals reaching Bob $n_{\mathrm{noise},B} = P_{\mathrm{int}}/(h f B_s)$, with $h$ Planck’s constant, $f$ the optical frequency, and $B_s$ the quantum signal bandwidth. The excess noise at Bob is $\xi_{B} \simeq 2 n_{\mathrm{noise},B}$, and referring this to the channel input gives $\xi_{\mathrm{coex}} = \xi_{B}/T = 2 P_{\mathrm{int}}/(T\,h\,f\,B_s)$, the equivalent number of additional vacuum units at the channel input.

For Gaussian‑modulated coherent‑state CV‑QKD with reverse reconciliation under collective attacks, the asymptotic SKR per channel use is $K = \beta I_{AB} - \chi_{BE}$, where $\beta$ is the reconciliation efficiency, $I_{AB}$ mutual information between Alice and Bob, and $\chi_{BE}$ the Holevo bound from the equivalent Gaussian channel~\cite{PhysRevResearch.6.023321}. The total input‑referred noise is $\chi_{\mathrm{tot}} = \chi_{\mathrm{line}} + \chi_{\mathrm{det}} + \xi_{\mathrm{coex}}$, where $\chi_{\mathrm{line}}$ is loss‑induced vacuum noise, $\chi_{\mathrm{det}}$ trusted detection noise (homodyne efficiency $\eta_B = 0.6$, electronic noise variance $V_{\mathrm{el}} = 0.01$) and $\xi_{\mathrm{coex}}$ coexistence noise from classical channels; in security analysis, $\chi_{\mathrm{line}} + \xi_{\mathrm{coex}}$ are treated as untrusted, while $\chi_{\mathrm{det}}$ is trusted.

The coexistence model solves coupled power‑evolution equations for SpRS and FWM along the fibre. FWM‑induced noise scales as $\gamma^{2} P^{3} L_{\mathrm{eff}}^{2}$ and decays with frequency separation $\Delta f$ through a phase‑matching factor $\propto 1/\!\left[1 + (\Delta\beta L_{\mathrm{eff}})^{2}\right]$, where $\Delta\beta \propto \beta_{2} \Delta f^{2}$ is the phase mismatch; here, $\gamma$, $L_{\mathrm{eff}}$, and $\beta_{2}$ denote the nonlinear coefficient, effective fibre length, and dispersion parameter, respectively. Thus, widening spectral guardbands efficiently suppresses FWM~\cite{PhysRevApplied.14.024013, agrawal2007nonlinear}. In contrast, SpRS scales linearly as $g_{R} P L_{\mathrm{eff}}$ (with Raman coefficient $g_R$) across a broad bandwidth spanning $\sim 40$~THz, contributing noise throughout the C-band and making it largely insensitive to gigahertz-scale guardbands~\cite{agrawal2007nonlinear}. For the 88\,$\times$\,50~GHz C‑band grid, these dependencies produce a transition region for per‑channel powers between $-5$ and $0$~dBm where FWM overtakes SpRS as the dominant coexistence noise, motivating guardband optimisation.

\vspace{-0.2cm}
\begin{figure}[]
    \vspace{-0.1cm}
    \centering
    \input{tikz/Figure.tex}
    \vspace{-10pt}
    \caption{(a) Spectral FWM and SpRS contributions at $-4.5$~dBm/ch. (b) Guardband effect across powers: 0.5, $-1.5$, $-4.5$~dBm/ch. (c) SKR–capacity tradeoff at $-1.5$~dBm/ch for co‑ and counter‑propagation. (d) Distance reach with/without 3‑channel guardbands.}
    \vspace{-2mm}
    \label{fig:combined}
\end{figure}

\vspace{-2mm}
\section{Numerical Analysis and Guardband Optimisation}
\vspace{-2mm}
We sweep guardband size $N_{\text{GB}}$ (0–10 channels), per-channel launch power ($-4.5$ to $0.5$~dBm/ch) and distance (0–30~km) on an $N_{\mathrm{cl}} = 88$ channel, 50~GHz DWDM grid. The quantum channel is placed at Ch~44 (band centre) or Ch~88 (band edge), resulting in classical capacity losses of $\Delta C = 2N_{\text{GB}}/88$ and $N_{\text{GB}}/88$, respectively. 

Figure~\ref{fig:combined}(a) separates FWM and SpRS contributions at $-4.5$~dBm/ch: FWM noise is concentrated near the band centre, while SpRS varies by $<10\%$ across the grid, so placing the quantum channel at the band edge minimises the impact of both mechanisms. Using Ch~88 as representative, Fig.~\ref{fig:combined}(b) shows a clear power dependence of the guardband benefit: at high power ($0.5$~dBm/ch), FWM dominates and a 3-channel (150~GHz) guardband recovers SKR from $\sim 0$ to 38~Mbit/s, whereas at low power ($-4.5$~dBm/ch) SpRS dominates and SKR stays around 200~Mbit/s independently of $N_{\text{GB}}$. At intermediate power ($-1.5$~dBm/ch), a 3-channel guardband yields a $108\%$ increase in SKR, indicating a transition to the FWM-limited regime. 

Figure~\ref{fig:combined}(c) shows the deployment rule: at $-1.5$~dBm/ch, band-edge placement with a 3-channel guardband achieves the $108\%$ SKR gain for only $3.4\%$ capacity loss, compared to $6.8\%$ for symmetric centre-band protection, so allocating guardband channels at the edge is twice as efficient. Counter-propagation provides only a marginal ($\sim 2.6\%$) additional SKR gain, suggesting that simple co-propagation with guardbands is already close to optimal. In terms of reach, Fig.~\ref{fig:combined}(d) shows that at high power ($0.5$~dBm/ch) a 3-channel guardband extends the co-propagating distance from $\sim 3$ to $\sim 10$~km. In contrast, at low power ($-4.5$~dBm/ch) all configurations reach 20–23~km, confirming that guardbands are only needed in the FWM-dominated power range.

\vspace{-2mm}
\section{Conclusion}
\vspace{-2mm}
Metropolitan CV-QKD coexistence with classical traffic is achievable when guardband optimisation is guided by the transition between FWM- and SpRS-dominated regimes. These theoretically derived guidelines identify where each impairment dominates to maximise efficiency. For powers below $-2$~dBm, the link is SpRS-limited and no guardband is required, whereas in the FWM-dominated regime ($-1.5$ to $0.5$~dBm/ch), band-edge placement with 150~GHz guardbands doubles SKR while incurring only a $3.4\%$ capacity loss, about half that of symmetric band-centre. These analytical results provide practical rules for deploying hybrid networks on existing fibre infrastructure.

\footnotesize
This work was supported in part by the European Union’s Horizon Europe Research and Innovation Programme QuNEST Project under the Marie Sklodowska-Curie Grant Agreement No. 101120422
\vspace{-0.4cm}
\footnotesize

\bibliographystyle{opticajnl}
\bibliography{short.bib} 

\end{document}

%% file: tikz/Figure.tex
\begin{tikzpicture}
    \begin{axis}[
        width=7.5cm,
        height=5cm,
        xshift=-1.5cm,
        xlabel={DWDM Channel Number},
        xlabel near ticks,
        xlabel shift=-3.0pt,
        ylabel={SKR at 10 km (Mbit/s)},      
        ylabel near ticks,
        ylabel shift=-3.0pt,
        grid=major,
        grid style={dashed, gray!30},
        label style={scale=0.85},
        tick label style={scale=0.85},
        xmin=1,
        xmax=88,
        ymin=0,
        ymax=170,
        enlargelimits=false]

    \addplot[C0, solid, very thick, mark=none] 
        table[x expr=\thisrow{Channel}+1, y expr={(\thisrow{SKR_bits_symbol}*5e8)/1e6}, col sep=comma] 
        {Data_Files_2/15dBm_FWM_Only_10km.csv};
    
    \addplot[C3, dashed, very thick, mark=none] 
        table[x expr=\thisrow{Channel}+1, y expr={(\thisrow{SKR_bits_symbol}*5e8)/1e6}, col sep=comma] 
        {Data_Files_2/15dBm_SpRS_Only_10km.csv};
    
    \addplot[black, solid, very thick, mark=none] 
        table[x expr=\thisrow{Channel}+1, y expr={(\thisrow{SKR_bits_symbol}*5e8)/1e6}, col sep=comma] 
        {Data_Files_2/15dBm_Combined_10km.csv};
    
    \fill[C2l, opacity=0.2] (axis cs:1,0) rectangle (axis cs:6,\pgfkeysvalueof{/pgfplots/ymax});
    \fill[C2l, opacity=0.2] (axis cs:83,0) rectangle (axis cs:88,\pgfkeysvalueof{/pgfplots/ymax});

    \node[C0, font=\scriptsize] at (axis cs: 44.5, 141) {FWM-Only};
    \node[C3, font=\scriptsize] at (axis cs: 44.5, 100) {SpRS-Only};
    \node[black, font=\scriptsize] at (axis cs: 44.5, 65) {FWM+SpRS};

    \node[C2, rotate = 90, font=\scriptsize] at (axis cs: 3.5, 30){Band-Edge};
    \node[C2, rotate = 90, font=\scriptsize] at (axis cs: 85.5, 30){Band-Edge};

    \node[font=\footnotesize, anchor=north west] at (rel axis cs: 0, 1) {(a)};
    \end{axis}

    \begin{axis}[
        xshift=7cm,
        width=7.5cm,
        height=5cm,
        xlabel={Guardband Size (Channels)},
        xlabel near ticks,
        xlabel shift=-3.0pt,
        ylabel={SKR at 10 km (Mbit/s)},
        ylabel near ticks,
        ylabel shift=-3.0pt,
        grid=major,
        grid style={dashed, gray!30},
        label style={scale=0.85},
        tick label style={scale=0.85},
        xmin=0,
        xmax=10,
        ymode=log,
        ymin=1e0,
        ymax=2e2,
        enlargelimits=false,
        legend style={
            draw=black,
            fill=white,
            at={(0.80, 0.01)},
            anchor=south,
            legend columns=1,
            font=\scriptsize,
            column sep=2pt,
            row sep=0.01pt,
            inner xsep=0.01pt,
            inner ysep=0.2pt,
            cells={anchor=west},
        },
        legend image post style={sharp plot, mark repeat=2},
        set layers, mark layer=axis tick labels
    ]
    grid=major,


    \addplot[C0, solid, very thick, mark=*, mark size=1.5pt] 
    table[col sep=comma, x=QSpace, y expr={(\thisrow{SKR_10km}*5e8/1e6)}]
    {Data_Files_2/Ppower_20dBm_Qch_87.csv};
    \addlegendentry{0.5 dBm/ch}
    \draw[dotted, thick] (axis cs: 0, 2.7e1) -- (axis cs: 2.0,2.7e1);

    \addplot[C2, solid, very thick, mark=*, mark size=1.5pt] 
    table[col sep=comma, x=QSpace, y expr={(\thisrow{SKR_10km}*5e8/1e6)}]
    {Data_Files_2/Power_18dBm_Qch_87.csv};
    \addlegendentry{-1.5 dBm/ch}

    
    
    \addplot[C3, solid, very thick, mark=*, mark size=1.5pt] 
    table[col sep=comma, x=QSpace, y expr={(\thisrow{SKR_10km}*5e8/1e6)}]
    {Data_Files_2/Ppower_15dBm_Qch_87.csv};
    \addlegendentry{-4.5 dBm/ch}

   \draw[{Latex[length = 1mm]}-{Latex[length = 1mm]}, black!80, thick] (axis cs: 2.0, 2.7e1) -- (axis cs: 2.0, 6e1);
   \node[black!80, font=\scriptsize, below right] at (axis cs: 1.9, 5.2e1) {108\% gain};
   

    \node[font=\footnotesize, anchor=north west] at (rel axis cs: 0, 1) {(b)};
    \end{axis}

    \begin{axis}[
        yshift=-4.5cm,
        xshift=-1.5cm,
        width=7.5cm,
        height=5cm,
        xlabel={Guardband Size (Channels)},
        xlabel near ticks,
        xlabel shift=-3.0pt,
        ylabel={\shortstack{SKR at 10 km (Mbit/s)}},
        ylabel style={color=C0},
        ylabel near ticks,      
        ylabel shift=-3.0pt,
        axis y line*=left,
        grid=major,
        grid style={dashed, gray!30},
        label style={scale=0.85},
        tick label style={scale=0.85},
        xmin=0,
        xmax=10,
        enlargelimits=false,
        legend style={
            draw=black,
            fill=white,
            at={(0.79, 0.01)},
            anchor=south,
            legend columns=1,
            font=\scriptsize,
            column sep=2pt,
            row sep=0.01pt,
            inner xsep=0.01pt,
            inner ysep=0.2pt,
            cells={anchor=west},
        },
        legend image post style={sharp plot}
    ]

\addplot[color=C0, solid, very thick, mark=*, mark size=1.5pt] 
    table[col sep=comma, x=QSpace, y expr={(\thisrow{SKR_bits_symbol}*5e8)/1e6}] 
    {Data_Files_2/2nd10km_tradeoff_18dBm87thCo.csv};
    \addlegendentry{Co-prop.}

\addplot[color=C0, solid, very thick, mark=triangle*, mark size=1.8pt] 
    table[col sep=comma, x=QSpace, y expr={(\thisrow{SKR_bits_symbol}*5e8)/1e6}] 
    {Data_Files_2/2nd10km_tradeoff_18dBm87thCounter.csv};
      \addlegendentry{Counter-prop.}

\node[font=\footnotesize, anchor=north west] at (rel axis cs: 0, 1) {(c)};
\end{axis}

    \begin{axis}[
        yshift=-4.5cm,
        xshift=-1.5cm,
        width=7.5cm,
        height=5cm,
        axis y line*=right,
        axis x line=none,
        ylabel={\shortstack{Classical Capacity Loss (\%)}},
        xmin=0, xmax=10,
        ymin=0, ymax=25,
        ylabel near ticks,      
        ylabel shift=-3.0pt,
        y label style={scale=0.85},
        y tick label style={scale=0.85},
    ]

\draw[black!80, thick]
    (axis cs: 3.2, 5.5) ellipse[y radius=10pt, x radius=4pt];
\draw[-
{Latex}, black!80, thick] (axis cs: 3.2, 3.0) -- (axis cs: 4.2, 3.0);

\draw[black!80, thick]
    (axis cs: 2.6, 20) ellipse[y radius=8pt, x radius=3pt];
\draw[-{Latex}, black!80, thick] (axis cs: 2.6, 22) -- (axis cs: 1.6, 22);
    
\addplot[color=C0, dashdotted, very thick, forget plot] 
    table[col sep=comma, x=QSpace, y=CapacityLoss] 
    {Data_Files/10km_tradeoff_15dBm87thCo.csv};

\addplot[color=C3, dashdotted, very thick, forget plot] 
    table[col sep=comma, x=QSpace, y=CapacityLoss] 
    {Data_Files/10km_tradeoff_15dBm44thCo.csv};

\node[C3, font=\scriptsize, anchor=west, rotate=28] at (axis cs: 5.0, 10.0) {Band-Center};
\node[C0, font=\scriptsize, anchor=west, rotate=17] at (axis cs: 5.3, 7.0) {Band-Edge};

    \end{axis}

\begin{axis}[
    yshift=-4.5cm,
    xshift=7cm,
    width=7.5cm,
    height=5cm,
    xlabel={Distance (km)},
    xlabel near ticks,
    ylabel={SKR (Mbit/s)},
    ylabel near ticks,
    ymode=log,                      
    xmin=1, xmax=30,
    ymin=5E-2, ymax=9e4,
    grid=major,
    grid style={dashed, gray!30},
    label style={scale=0.85},
    tick label style={scale=0.85},
    legend style={
    draw=black,
    fill=white,
    at={(0.98, 0.98)},
    anchor=north east,
    font=\scriptsize,
    row sep=0pt,
    inner xsep=1pt,
    inner ysep=1pt,
    cells={anchor=west},
},
legend image post style={sharp plot, xscale=0.5}
]

\addlegendimage{C0, solid, very thick}
\addlegendentry{0 ch, Co-prop.}

\addlegendimage{C0, dashed, very thick}
\addlegendentry{0 ch, Counter-prop.}

\addlegendimage{C3, solid, very thick}
\addlegendentry{3 ch, Co-prop.}

\addlegendimage{C3, dashed, very thick}
\addlegendentry{3 ch, Counter-prop.}

\draw[black!80, thick] (axis cs: 24,5e0) arc(0:180:12pt and 3pt); 
\draw[black!80, thick] (axis cs: 11.0, 1e1) arc(0:180:27pt and 3pt);

\addplot[C0, solid, very thick] table[x=z, y expr={(\thisrow{SKR-Fw}*5e8)/1e6+1e-5}, col sep=comma] {Data_Files/TotalInterference/QSpace-0/Power_15dBm_Qch_87.csv};
\addplot[C0, solid, very thick] table[x=z, y expr={(\thisrow{SKR-Fw}*5e8)/1e6+1e-5}, col sep=comma] {Data_Files/TotalInterference/QSpace-0/Power_20dBm_Qch_87.csv};

\addplot[C0, dashed, very thick] table[x=z, y expr={(\thisrow{SKR-Bw}*5e8)/1e6+1e-5}, col sep=comma] {Data_Files/TotalInterference/QSpace-0/Power_15dBm_Qch_87.csv};
\addplot[C0, dashed, very thick] table[x=z, y expr={(\thisrow{SKR-Bw}*5e8)/1e6+1e-5}, col sep=comma] {Data_Files/TotalInterference/QSpace-0/Power_20dBm_Qch_87.csv};

\addplot[C3, solid, very thick] table[x=z, y expr={(\thisrow{SKR-Fw}*5e8)/1e6+1e-5}, col sep=comma] {Data_Files/TotalInterference/QSpace-3/Power_15dBm_Qch_87.csv};
\addplot[C3, solid, very thick] table[x=z, y expr={(\thisrow{SKR-Fw}*5e8)/1e6+1e-5}, col sep=comma] {Data_Files/TotalInterference/QSpace-3/Power_20dBm_Qch_87.csv};
    
\addplot[C3, dashed, very thick] table[x=z, y expr={(\thisrow{SKR-Bw}*5e8)/1e6+1e-5}, col sep=comma] {Data_Files/TotalInterference/QSpace-3/Power_15dBm_Qch_87.csv};
\addplot[C3, dashed, very thick] table[x=z, y expr={(\thisrow{SKR-Bw}*5e8)/1e6+1e-5}, col sep=comma] {Data_Files/TotalInterference/QSpace-3/Power_20dBm_Qch_87.csv};

\draw[black!80, thick] (axis cs: 24,5e0) arc(360:180:12pt and 3pt);

\node[black!80, font=\scriptsize, below left] at (axis cs: 19.7, 1.0e1) {-4.5};
\node[black!80, font=\scriptsize, below left] at (axis cs: 20.9, 4e0) {dBm/ch};

\draw[black!80, thick] (axis cs: 11.0, 1e1) arc(360:180:27pt and 3pt);
\node[black!80, font=\scriptsize, below right] at (axis cs: 5.3, 7.e0) {0.5};
\node[black!80, font=\scriptsize, below right] at (axis cs: 4.3, 3.0e0) {dBm/ch};

\node[font=\footnotesize, anchor=north west] at (rel axis cs:0,1) {(d)};
\end{axis}

\end{tikzpicture}